\documentclass[12pt]{iopart} 
\usepackage{amsmath}
\usepackage{amssymb}
\usepackage{graphicx}
\usepackage{upgreek}
\usepackage{hyperref}
\usepackage{xcolor}
\usepackage{orcidlink}

\def\df#1{{\sf#1}}
\def\be{\begin{equation}}
\def\ee{\end{equation}}
\def\bea{\begin{eqnarray}}
\def\eea{\end{eqnarray}}
\def\revision#1{{#1}}

\begin{document}
\bibliographystyle{iopart-num}

\title[On the possibility of classical vacuum polarization and magnetization]{On the possibility of classical vacuum polarization and magnetization}

\author{S\'ebastien Fumeron$^a$, Fernando Moraes$^b$ \orcidlink{0000-0001-7045-054X}, Bertrand Berche$^{a,c}$  \orcidlink{0000-0002-4254-807X}}

\address{$^a$ Laboratoire de Physique et Chimie Th\'eoriques,
 Universit\'e de Lorraine - CNRS, Nancy, France\\
 $^b$ Departamento de F\'{\i}sica, Universidade Federal Rural de Pernambuco, 
52171-900, Recife, PE, Brazil\\
$^c$ ${\mathbb L}^4$ Collaboration \& Doctoral College for the Statistical Physics of Complex Systems, Leipzig-Lorraine-Lviv-Coventry, Europe
}
\ead{sebastien.fumeron@univ-lorraine.fr, fernando.jsmoraes@ufrpe.br, bertrand.berche@univ-lorraine.fr}
\vspace{10pt}
\begin{indented}
\item[] \today
\end{indented}

%
%
%

\begin{abstract}
     It is common practice to take for granted the equality (up to the constant $\varepsilon_0$) of the electric displacement ($\bf{D}$) and electric ($\bf{E}$) field vectors in vacuum. The same happens with the magnetic field ($\bf{H}$) and the magnetic flux density ($\bf{B}$) vectors (up to the constant $\mu_0^{-1}$).  The fact that gravity may change this by effectively inducing dielectric or magnetic responses to the primary fields is commonly overlooked. It is the purpose of this communication to call attention to  classical polarization or magnetization of the vacuum due to the concomitant presence of  gravitational and electromagnetic sources. The formalism of differential forms (exterior calculus) is used since it provides a clear-cut way to achieve this. \revision{This work  offers new routes for possible detection of various spacetime geometries via their electromagnetic manifestations and the way they influence light propagation.}
\end{abstract}

\maketitle

\submitto{\CQG}

\section{Introduction}

Vacuum polarization is a well-identified phenomenon in quantum electrodynamics. Since the pioneering works of  Dirac \cite{dirac1934discussion}, Furry and Oppenheimer \cite{Furry1934theory}, Heisenberg \cite{heisenberg1989bemerkungen1}, Uehling \cite{uehling1935polarization} and Weisskopf \cite{weisskopf1936electrodynamics}, vacuum is understood as a dynamical object filled with quantum fluctuations. As prescribed by the Heisenberg indeterminacy relations, virtual electron-positron pairs can indeed briefly pop in and out of existence to interact with the external electromagnetic field (EM field), in an analog fashion to what happens inside any polarizable medium: vacuum permittivity value $\varepsilon_0=8.854\ \!\! 187\ \!\! 82\ 10^{-12}\:\hbox{F.m}^{-1}$ corresponds to the particular case for which vacuum is maximally polarized \cite{leuchs2020qed}. 

Quantum vacuum polarization manifests itself in a large variety of situations, including the Casimir effect, the Hawking radiation, and the Lamb shift. 
In contrast, the possibility of a classical vacuum polarization is less often (if almost ever) considered in the literature \cite{cabral2017electrodynamics,cabral2017electrodynamicsII,PhysRevD.105.105026}. It can be defined as {\em any deviation of the electric constitutive relation from the form it takes in the flat Minkowski spacetime}. In this paper, we will use exterior calculus to investigate the possibility of vacuum polarization and magnetization in curved spacetimes containing electromagnetic sources. 

In this formalism, and considering units such that $\varepsilon_0=\mu_0=1$, Maxwell's equations in vacuum reduce to
 {Bianchi equation} $d\df F=0$  {for the Faraday $2-$form defined in terms of the potential $1-$form, $\df F= d\df A$}, and outside the location of point charges, $d\df G=0$. Here $\df F=\df E\wedge dt + \df B$  and $\df G=\df D-\df H\wedge dt$ is the Maxwell 2-form (differential forms are denoted in sanserif to distinguish them from their components in italics)\cite{info}. These are geometry-independent equations but the constitutive relation $\df G=\star_4 \df F$ on the other hand \revision{(see Appendix for a short introduction to the ``star operation'')},  which follows from the definition of an action in the form of
\be \revision{S[\df A]=\int \frac{1}{2}\df F\wedge\star_4\df F-\df A\wedge\star_4\df J,}\label{Eq-action}\ee
does depend on the geometry, which is embedded in the Hodge star operator. This is the cause of the classical  polarization and magnetization of the vacuum in local coordinates.  
For a recent introduction to exterior calculus applied to electrodynamics, including classical and quantum vacuum polarization, see Ref. \cite{fumeron2020improving}.

In order to have a visual impression of polarization or magnetization effects on the electromagnetic fields, for the various geometries considered, we \revision{display} plots of the field lines \revision{as if}  the expressions found \revision{in terms of local coordinates were those} in flat spacetime. Of course, this distorts the field lines, since they are in reality in curved spacetime, but \revision{preserves}  their topology.  

The paper is structured as follows. First, the electrostatics of Reissner-Nordström (RN) and related spacetimes will be studied in Section \ref{RNsection}. In Subsection \ref{RNsubsection} the RN spacetime 
is used as groundwork to determine a general condition for vacuum polarization to arise. Then, in Subsection \ref{stringRNsection} we will establish how the additional presence of a cosmic string may indeed couple to the EM field such as to produce a non-trivial polarization in RN spacetime. Classical vacuum polarization  will also be found in the case of a charged wormhole (Subsection \ref{chargedwormsection}). In a similar way, classical  magnetization is studied in Section \ref{Melvinsection} with its subsections focusing on Melvin  (Subsection \ref{Melvinsubsection}) and Ernst  (Subsection \ref{ernstsection}) spacetimes. The case of a rotating charged gravitational source, the Kerr-Newman (KN) spacetime, where both polarization and magnetization appear, is studied in Section \ref{KNsection}. Finally, in Section  \ref{conclsection} we  will present our conclusions.

\section{Electrostatics in Reissner-Nordström 
spacetimes \label{RNsection} }

\subsection{RN spacetime \label{RNsubsection}}
The very first exact solution of Einstein's field equations was found in 1916 by Karl Schwarzschild \cite{Schw1916}. This is the so-called Schwarzschild metric which describes the geometry of spacetime in the vicinity of a static and spherically-symmetric compact source of gravitation such as a star or a black hole. Soon after, Weyl, Reissner and Nordstr\"{o}m independently considered a generalization of the Schwarzschild solution when the compact object (black hole) has a net charge $Q$ in addition to the mass parameter $M$. As found by Bekenstein \cite{Bekenstein} in 1971, the  gravitational field near a charged star is the standard Reissner-Weyl-Nordstr\"{o}m  metric as well. 

In standard units where $c=1$ and $G=1$, the Reissner-Nordström metric line element in local coordinates $(t,r,\theta,\varphi)$ writes  as
\begin{eqnarray}
g&=&-\left(1-\frac{2M}{r}+\frac{Q^2}{r^2}\right)dt^2+\frac{dr^2}{1-\frac{2M}{r}+\frac{Q^2}{r^2}} \nonumber \\
&&\quad+r^2\left(d\theta^2+\sin^2\theta\: d\varphi^2\right),
\label{metric}
\end{eqnarray}
for $r>R$. Here, $R$, $M$ and $Q$ represent the radius, mass and charge of the star, respectively. 
The metric line element, written above in the coordinate basis,  takes the standard Minkowskian form
$g=-(\df e^0)^2+(\df e^1)^2+(\df e^2)^2+(\df e^3)^2$
 in the local coframe $\df e^0=\sqrt{A(r)}\:dt$, $\df e^1=dr/\sqrt{A(r)}$, $\df e^2=rd\theta$ and $\df e^3= r\sin\theta d\varphi$ with 
$A(r)= 1-\frac{2M}{r}+\frac{Q^2}{r^2}$.
 There, spherical symmetry
demands that the 2-form $\df D$ has the simple expression $\df D=D_{23}(r)\df e^2\wedge \df e^3$.
Since the total charge is $Q$, 
and $\int_{\partial\mathcal{V}} \df D = Q$ \revision{from Gauss theorem $d\df D=\rho$ with $\rho$ the charge density $3-$form}, it turns out that the electric flux density 2-form is given by a single component in the coordinate basis
\begin{equation}
    \df D= \frac{Q}{4\pi r^2}e^2\wedge \df e^3=
 \frac{Q}{4\pi} \sin\theta \:d\theta \wedge d\varphi, \label{Gauss}
\end{equation}
where we read that $D_{\theta\varphi}= \frac{Q}{4\pi} \sin\theta$.
The electric field 1-form $\df E$ is obtained from the Hodge star operator as
\begin{equation}
 \df    D=\star_4 \left(\df E \wedge dt\right) \label{DfromE}
\end{equation}
\revision{where $\star_4$ is the Hodge dual operator (see Appendix), which, applied to any $p-$form
$\df u$, completes $\df u$ to the $4-$volume form, $\df u\wedge\star_4\df u=\frac 1{p!}u_{\mu_1...\mu_p}u^{\mu_1...\mu_p} \sqrt{-\hbox{det}\ \!g}\ \! dx^{\mu_1}\wedge...dx^{\mu_4}$. This is the key property which enables one to construct actions like in equation~(\ref{Eq-action}).}

Straightforward algebra shows that
\begin{eqnarray}
  \star_4\left(d\theta \wedge d\varphi\right)=\frac{1}{r^2\sin\theta} dt \wedge dr
\end{eqnarray}
so that 
\begin{equation}
    \star_4\df D=-\df E \wedge dt=\frac{Q}{4\pi r^2}dt \wedge dr 
\end{equation}
Hence, the unique component of the 1-form $\df E$ in the coordinate basis is equal to 
\begin{equation}
    E_r=\frac{Q}{4\pi r^2}=\frac{1}{r^2\sin\theta}D_{\theta\varphi} \label{EfieldRN}
\end{equation}
This means that the vacuum polarization due to the Reissner-Nordström spacetime, with permittivity $\varepsilon_r(r,\theta)=r^2\sin\theta$,
is exactly the same as in ordinary  empty space. The result  $\varepsilon_r(r,\theta)=r^2\sin\theta$ is rather a manifestation of the local spherical coordinates than a true vacuum polarization.

\subsection{RN spacetime pierced by a cosmic string \label{stringRNsection}}

In order to investigate a situation that does not reduce to empty \revision{flat} space \revision{at infinity}, we can consider the case of a Nambu-Goto cosmic string, with infinite length and zero thickness. 
Cosmic strings are topological defects associated with a conical geometry obtained by cutting a wedge in the background spacetime\revision{\cite{Vilenkin:2000jqa,FB2023}}
\begin{eqnarray}
g=-dt^2+d\rho^2+\alpha^2 \rho^2d\varphi^2+dz^2 , \label{string}
\end{eqnarray}
(in local coordinates $(t,\rho,\varphi,z)$ with the usual meaning $\rho=r\sin\theta$ and $z=r\cos\theta$) where $0\leq \varphi < 2\pi$ and $\alpha=1-4\mu$. \revision{The string tension $\mu$ is related to the mass per unit length of the defect and in first approximation, it is estimated from $G\mu=(\eta/M_P)$ where $\eta$ is the energy scale of the string-forming phase transition and $M_P$ is the Planck mass. Comparison between simulations of the cosmic microwave background (CMB) in the presence of Nambu-Goto strings (unconnected segment model) and observational datas of the CMB power spectrum from Planck set a $95\%$ confidence upper limit of $G\mu < 1,5.10^{-7}$\cite{Planck2013}. Methods based on gravitational wave interferometry decreased the upper limit of several order of magnitude and currently, estimation of the string tension is narrowed down to $G\mu <4.10^{-15}$ \cite{auclair2023window}.}

\revision{In spite of the absence of experimental evidence of such a spacetime,} we now consider the case of a Reissner-Nordström black hole crossed by a cosmic string, \revision{in order to combine both the effect of the black hole and of the conical geometry which makes the metrics deviate from flat spacetime even at infinity}. The metric line element in $(t,r,\theta,\varphi$) coordinates and in standard units writes as  \cite{Linet1999}
\begin{eqnarray}
g&=&-\left(1-\frac{2M}{r}+\frac{Q^2}{r^2}\right)dt^2+\frac{dr^2}{1-\frac{2M}{r}+\frac{Q^2}{r^2}} \nonumber \\
&&+\quad r^2\left(d\theta^2+\alpha^2\sin^2\theta\: d\varphi^2\right),
\label{RNstring}
\end{eqnarray}
i.e. a slight modification $\varphi\to\alpha\varphi$ compared tp the previous case.
The result \eqref{Gauss} is slightly modified, hence 
$D_{\theta\varphi}=[Q/(4\pi)]\alpha\sin\theta$, and
the same line of reasoning as before leads to 
\begin{equation}
  \df   E = \frac{Q}{4\pi  r^2} dr ,\label{E_RNstring}
\end{equation}
or equivalently
\begin{equation}
    D_{\theta\varphi} = \alpha r^2\sin\theta E_r .
\end{equation}

The cosmic string tension couples to the electromagnetic field and produces a vacuum polarization which does not reduce to the use of local spherical coordinates. This is clearly a manifestation  of the cosmic string since the result still holds for $M=Q=0$, which turns the metric \eqref{RNstring} into \eqref{string}. Also, in the absence of the string ($\alpha=1$) the RN result \eqref{EfieldRN}  is recovered.

\subsection{ Charged wormhole spacetime \label{chargedwormsection}}

A combination of Morris-Thorne wormhole and Reissner-Nordstr\"om spacetimes, the charged wormhole solution, is described by the metric
  \cite{kim2001exact} 

\begin{eqnarray}
g&=&-\left(1+\frac{Q^{2}}{r^{2}}\right) d t^{2}+\left(1-\frac{b(r)}{r}+\frac{Q^{2}}{r^{2}}\right)^{-1} d r^{2}\nonumber \\
&&\quad +r^{2}\left(d \theta^{2}+\sin ^{2} \theta d \varphi^{2}\right),\label{EQ12}
\end{eqnarray}
where we choose the Morris and Thorne \cite{morris1988wormholes} shape function $b(r)=b_0^2/r$.   \revision{The spatial shape of the MT wormhole is specified by the function $b(r)$, thus its denomination. The parameter $b_0$  defines the smaller possible value of $r$, i.e. the wormhole throat. In the MT case ($Q=0$) the coordinate $r$ is problematic at $r=b_0$ for the factor multiplying $dr^2$ in equation~(\ref{EQ12}) vanishes.    Furthermore,  the MT wormhole is unstable but this may be resolved by adding exotic matter \cite{morris1988wormholes} or electric charge \cite{kim2001exact}. }
In this case, the condition $Q^2 < b_0^2$ is required \cite{kim2001exact}  to maintain the wormhole throat open.  \revision{This also solves the problem at $r=b_0$. }
The reader should be aware that this  is a rather extreme case since, in order to have an open mouth of radius around $1~\hbox{m}$, the wormhole would need to have a charge of the order of $3\times 10^{16}\, \hbox{C}$. More realistic charge values may be found for different shape functions but this would unnecessarily complicate this example.

From Gauss' law, we get the same result $D_{23}(r)=\frac{Q}{4\pi r^2}$, hence the same  \eqref{Gauss}, namely,
\begin{equation}
    D_{\theta\varphi}=\frac{Q}{4\pi }\sin\theta. \label{Dworm}
\end{equation}
Using \eqref{Dworm}, \eqref{DfromE} and \eqref{Hodge-def}, we get the electric field 1-form single component 
\begin{equation}
    \df E=
     \sqrt{\frac{1+\frac{Q^{2}}{r^{2}}}{1-\frac{b_0^2}{r^2}+\frac{Q^{2}}{r^{2}}}} D_{23}\:dr =E_rdr,
\end{equation}
in agreement with \cite{kim2001exact}. 

Note that, for $b_0=0$ the RN result \eqref{EfieldRN} is recovered. It follows that
\begin{eqnarray}
D_{\theta\varphi}= \sqrt{\frac{1-\frac{b_0^2}{r^2}+\frac{Q^{2}}{r^{2}}}{1+\frac{Q^{2}}{r^{2}}}} r^2\sin\theta E_r =\epsilon(r,\theta) E_r. \label{perm}
\end{eqnarray}
In this example, the vacuum polarization is due to the coupling of $b_0$ to $Q$ in the geometry. 
{Figure \ref{Wormfields} is a  plot of the effective permittivity of  the charged wormhole spacetime as given by Eq. \eqref{perm}. We call the reader's attention to the fact that as $r\rightarrow \sqrt{b_0^2 - Q^2}$ (or $r\rightarrow 1$ in the plot of Fig.1) the permittivity goes quickly to zero. This remarkable quality implies the total reflection of electromagnetic waves incident on the wormhole \cite{garcia2002zero,schwartz2003total}. On the other hand, recent research on near-zero permittivity metamaterials has revealed a number of exotic electromagnetic properties like, for instance, ``squeeze the electromagnetic wave and make it tunnel through a deep subwavelength channel with arbitrary shape'' \cite{liu2017manipulating}. This offers a perspective of building metamaterial-based analog models for further study of the charged wormhole spacetime. 
\begin{figure}[htp!]
    \centering 
      \includegraphics[width=0.75\columnwidth]{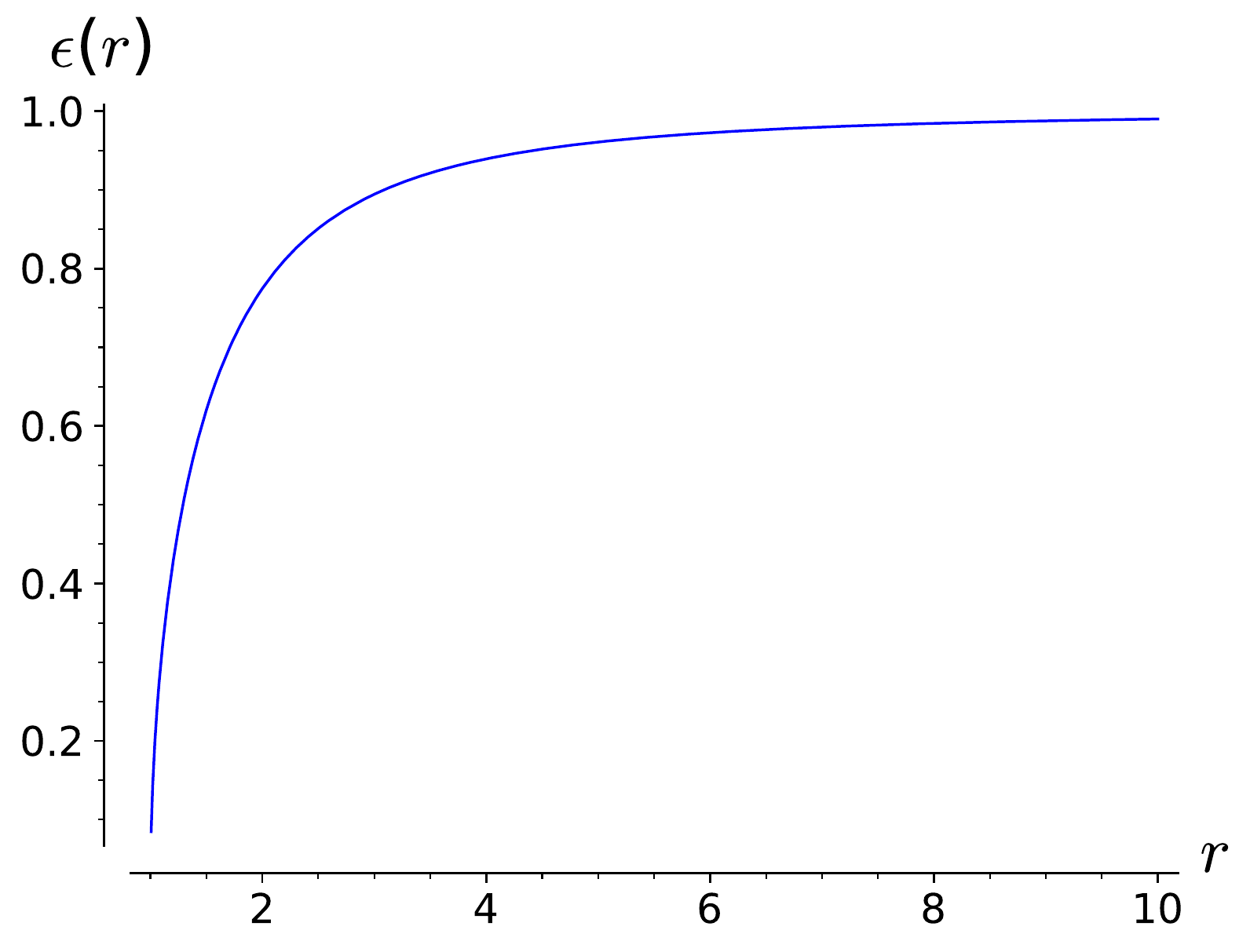} 
    \caption{Effective permittivity of the charged wormhole spacetime, in the equatorial plane ($\theta=\pi/2$). The parameters used for this plot were $b_0 =\sqrt{2}$ and $Q=1$. } 
   \label{Wormfields}
\end{figure}
Recalling that the wormhole connects two asymptotically flat spacetimes through a spherically symmetric bridge of radius $b_0$, the plot in Fig. \ref{Wormfields} represents the permittivity in either universe.  Since the range of the radial coordinate is $r \geq b_0 $, the permittivity does not really become zero but can reach arbitrarily small values depending on the relative values of $Q$ and $b_0$.  Again, quite suitable for a near-zero permittivity metamaterial analog model. 

We close this Section by noting that the robustness of the result (\ref{Dworm}) is a manifestation of the topological character of the relation $Q=\int_{\partial{\cal V}}\df D$ while the sensibility of the expression of $E_r$ with the form of the metric is a consequence of the use of Hodge duality. 

\section{Magnetostatics in Melvin and Ernst spacetimes \label{Melvinsection}}
\subsection{Melvin spacetime \label{Melvinsubsection}} 

The  Melvin magnetic universe is a solution of the Einstein-Maxwell equations associated with a bundle of magnetic flux lines held together by its own gravitational field \cite{melvin1964pure,melvin1965dynamics}. We note that there is also an electric solution analogous to this one, which can be obtained
by taking its electro-magnetic dual \cite{lim2018electric}. The line element of the magnetic spacetime is 
\begin{equation}
    g=-\Lambda(\rho)^{2} d t^{2}+\Lambda(\rho)^{2} d \rho^{2}+\Lambda(\rho)^{-2} \rho^{2} d \varphi^{2}+\Lambda(\rho)^{2} d z^{2}, \label{Melvinmetric}
\end{equation}
where 
\begin{equation}
    \Lambda(\rho)=1+\frac{1}{4}  \kappa_0^{2} \rho^{2} .
\end{equation}
Here, $ \kappa_0^{-1}$ is the  Melvin length scale, a  measure of the magnetic field strength $B_0$ on the axis, normalized to the dimensions of an inverse length $\kappa_0=B_0$.
The metric being diagonal, a simple tetrad choice reads as 
$\df e^0=\Lambda(\rho)dt$,  $\df e^1=\Lambda(\rho) dr$, $\df e^2=[\rho/\Lambda(\rho)]d\varphi$ and $\df e^3=\Lambda(\rho)dz$.

{An ansatz for the 1-form potential} in cylindrical coordinates is $\df A=A_2(\rho) \df e^2=A_2(\rho)[ {\rho}/{\Lambda(\rho)}]d\varphi$, therefore
 $A_2(\rho)$  follows from the definition of the dimensionless magnetic flux
$\Phi_0$ enclosed by a circle $\partial\Sigma$ of radius $\rho$ perpendicular to the $z$ axis,
\be
\int_{\partial\Sigma}\df A=\int_\Sigma \df B=\Phi_0=2\pi\rho A_2(\rho).\label{Eq-FluxCondition}
\ee
It follows that 
\be
A_2(\rho)=\frac{\Phi_0}{2\pi\rho},\quad\hbox{and}\quad A_\varphi(\rho)=\frac{\Phi_0}{2\pi}\frac{1}{\Lambda(\rho)}.\ee
This leads to
\be
\df F=\df B=d\df A=\partial_\rho A_\varphi d\rho\wedge d\varphi+\partial_zA_\varphi dz\wedge d\varphi
\ee
hence, 
\be B_{\varphi z}=0, \quad
B_{\rho\varphi}=-\frac{\Phi_0}{4\pi}\frac{\kappa_0^2\rho}{\Lambda(\rho)^2}.\ee
The 2-form $\df G$ is now given by the Hodge product $\df G=\star_4\df F=-\df H\wedge dt$. The calculation leads to
\be
\df G=-\frac{\Lambda(\rho)^2}{\rho}B_{\rho\varphi} dz\wedge dt
\ee
and implies 
\be
H_z=\frac{\Lambda(\rho)^2}{\rho}B_{\rho\varphi} =\frac{1}{\mu(\rho)}B_{\rho\varphi} 
\ee
with the relative permeability given by
\be
\mu(\rho)=\frac\rho{\Lambda(\rho)^2}.
\ee
Therefore there is  magnetization in Melvin spacetime since $H_z \neq B_{r\varphi}$. In Fig. \ref{Melvinfields} the magnetic fields are plotted in the $z-x$ plane assuming that the expressions for the fields are in a Minkowski background. 
\begin{figure}[htp!]
    \centering 
      \includegraphics[width=0.49\columnwidth]{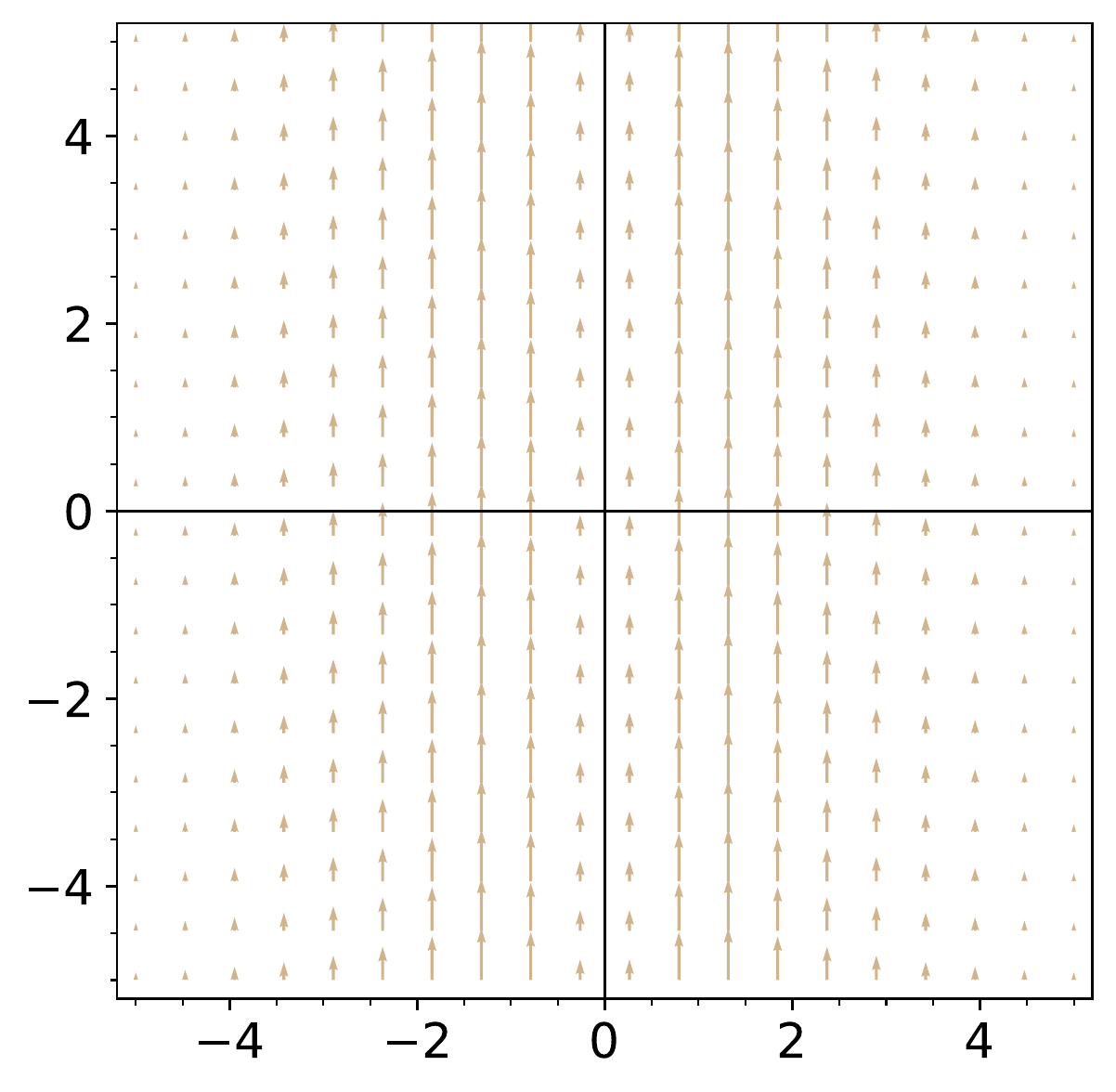} 
      \includegraphics[width=0.49\columnwidth]{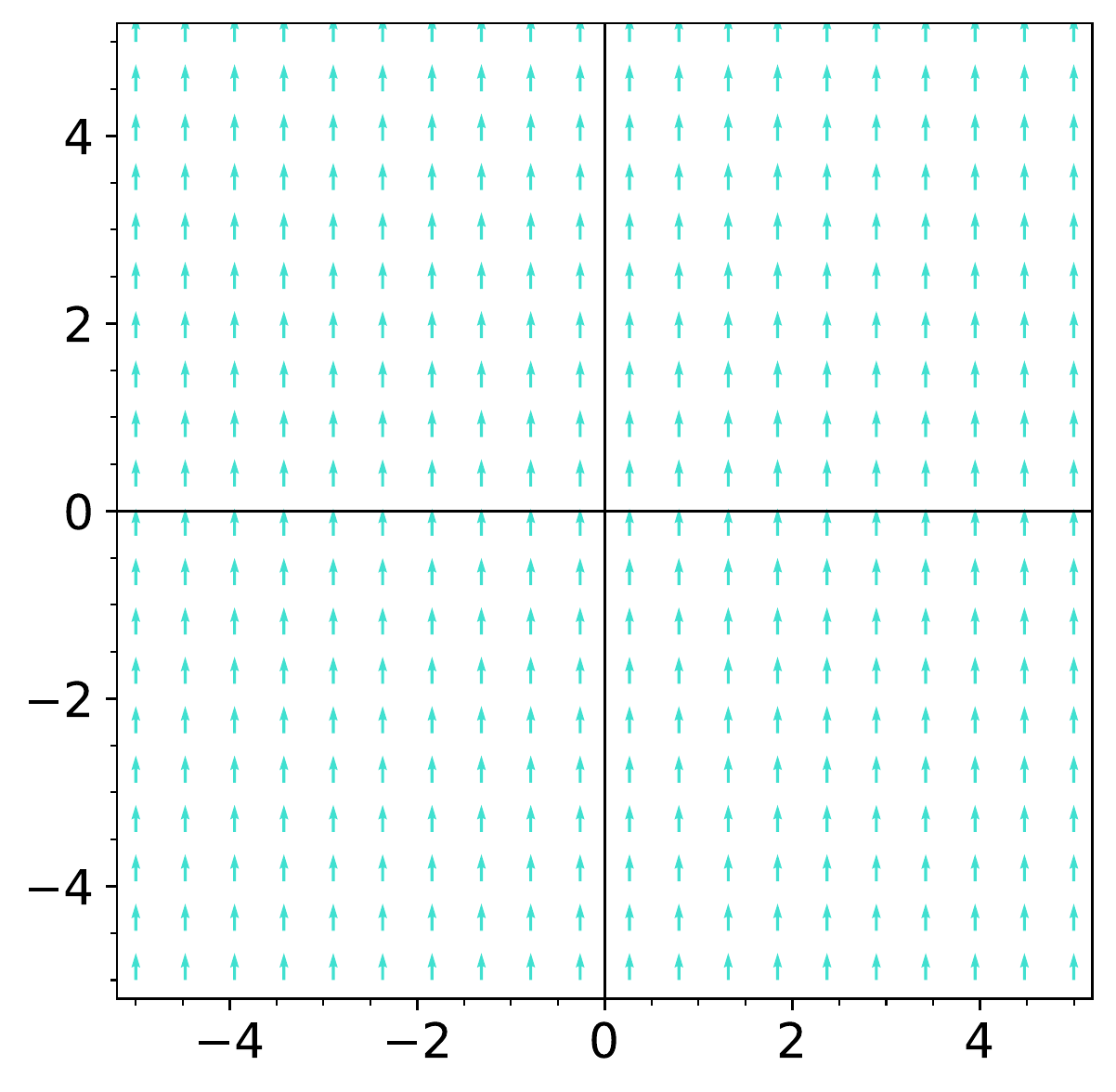} 
    \caption{Melvin magnetic flux density  $\mathbf{B}$ (left) and magnetic field $\mathbf{H}$ (right). The scale was set by choosing $ B_0=1$.}
   \label{Melvinfields}
\end{figure}

\subsection{Ernst spacetime \label{ernstsection}}

We now consider the case of an Ernst spacetime \cite{ernst1976black}, consisting of a black hole immersed in a magnetic field. The metric line element writes  in standard units as 
\begin{eqnarray}
g&=&\Lambda(r,\theta)^2\left[-\left(1-\frac{2M}{r}\right)dt^2 +\frac{dr^2}{1-\frac{2M}{r}}+r^2d\theta^2 \right] \nonumber \\
&&+\frac{r^2\sin^2\theta}{\Lambda(r,\theta)^2} d\varphi^2
\label{ernst}
\end{eqnarray}
 where $\Lambda(r,\theta)=1+\frac{\kappa_0^2}{4} r^2\sin^2\theta$ and $\kappa_0$ is, again, a constant normalized magnetic field. 
 
Then, one solves Maxwell's equations in the background metric (\ref{ernst}) with the 1-form potential $\df A$ obtained by the flux condition analogous to
 (\ref{Eq-FluxCondition}) \cite{Bytsenko2003}. 
 The coframe basis vectors follow from (\ref{ernst}) and read as
 \bea
&& \df e^0=\Lambda(r,\theta)\left(1-\frac{2M}{r}\right)^{1/2}dt,\\
&&  \df e^1=\Lambda(r,\theta)\left(1-\frac{2M}{r}\right)^{-1/2}dr,\\
&&  \df e^2 =\Lambda(r,\theta) rd\theta,\\ 
&&\df e^3=\frac{r\sin\theta}{\Lambda(r,\theta)}d\varphi
 \eea
 and assuming cylindrical symmetry in the Minkowski cotangent spacetime,
 $\df A=A_3(r\sin\theta)\df e^3$ we extract $A_3$ from
 \be
 \int\df A=\int d\df B=\Phi_0=2\pi r\sin\theta A_3(r,\theta).
 \ee
 It follows that\cite{footnoteBytsenko} 
 \be
 A_3(r,\theta)=\frac{\Phi_0}{2\pi r\sin\theta},\quad A_\varphi(r,\theta)=\frac{\Phi_0}{2\pi}\frac{1}{\Lambda(r,\theta)}
 \ee

Equation $\df F=\df B=d\df A$ \revision{then yields the magnetic $2-$form}
\begin{eqnarray}
\df B=-\frac{\Phi_0}{4\pi}\frac{\kappa_0^2 r\sin\theta}{\Lambda(r,\theta)^2}\left(\sin\theta\;dr\wedge d\varphi+r\cos\theta\;d\theta\wedge d\varphi\right)\nonumber\\
\label{F_E}
\end{eqnarray}
and from $\df G=\star_4\df F=-\df H\wedge dt$ one finds that 
$\df H$ expresses as
\begin{equation}
\df    H=\frac{\Lambda(r,\theta)^2}{\sin\theta}
\Bigl[\Bigl(1-\frac{2M}r\Bigr)B_{r\varphi} \ \!d\theta+\frac{1}{r^2}B_{\theta\varphi} \ \!dr\Bigr].
\end{equation}
\revision{This expression implies the relation between components}
\begin{eqnarray}
H_r&=&\frac{\Lambda(r,\theta)^2}{r^2\sin\theta} B_{\theta\varphi}, \\
H_{\theta}&=&\frac{\Lambda(r,\theta)^2}{\sin\theta}\left(1-\frac{2M}{r}\right) B_{r\varphi}.
\end{eqnarray}
This time, the background metric couples to the magnetic field, and, the metric  being not asymptotically flat  it results in an anisotropic magnetization. 

In Fig. \ref{Ernst_lines} the magnetic fields are plotted in the $z-x$ plane assuming, \revision{as in previous graphs,} that the expressions for the fields are in a Minkowski background. 

\begin{figure}[htp!]
    \centering 
      \includegraphics[width=0.46\columnwidth]{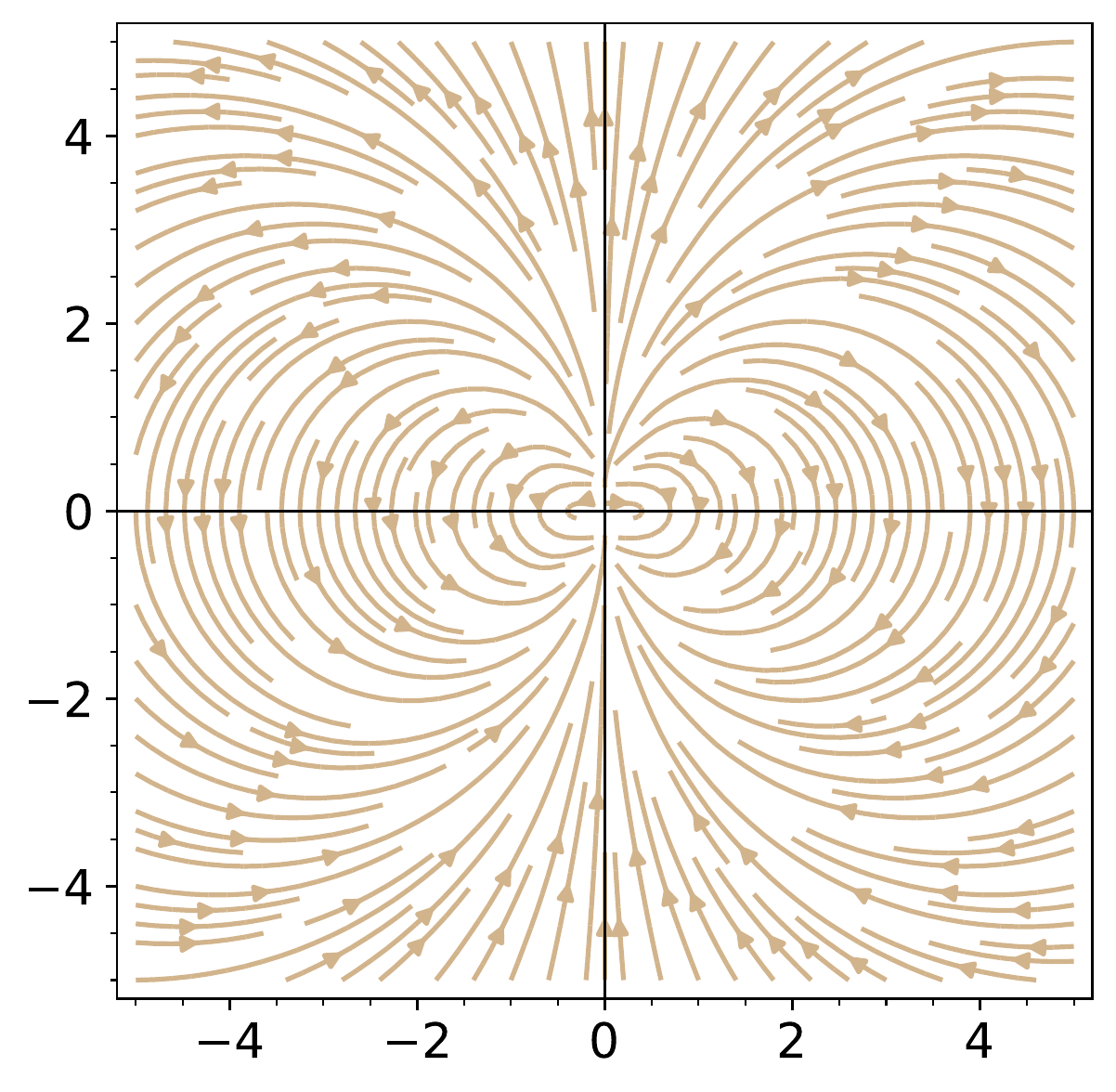} 
      \includegraphics[width=0.49\columnwidth]{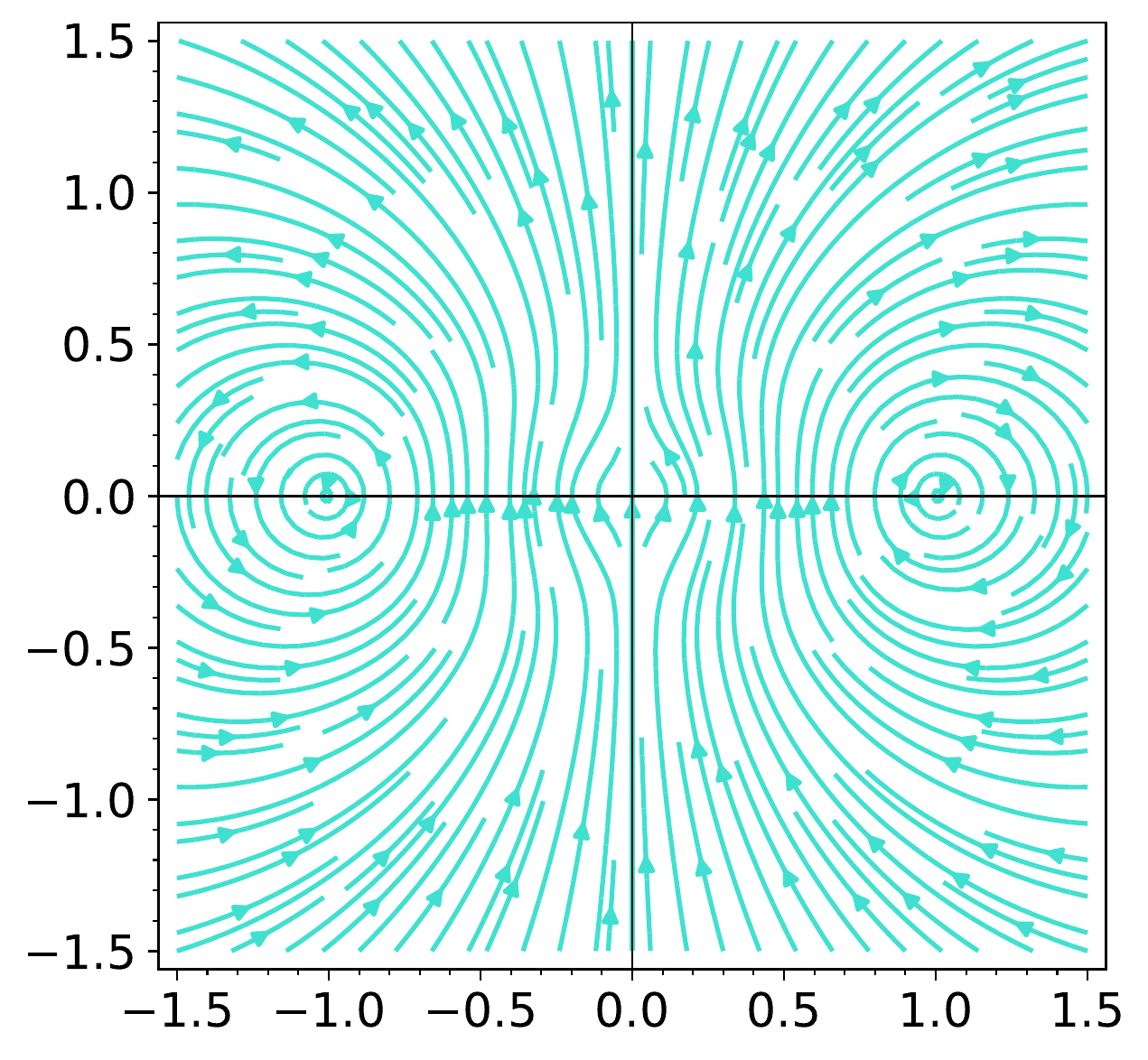} 
    \caption{Ernst magnetic flux density $\mathbf{B}$
 (left) and magnetic field $\mathbf{H}$ (right) lines. The scale was set by choosing $B_0=2M=1$.}
   \label{Ernst_lines}
\end{figure}

\section{Electric and magnetic fields in the Kerr-Newman spacetime \label{KNsection}}

As seen in Section \ref{RNsection}, the Reissner-Nordstr\"om solution   of Einstein-Maxwell equations generalizes the Schwarzschild spacetime to include charge of the gravitational source. Analogously, the Kerr solution \cite{kerr1963gravitational} complements Schwarzschild's by including rotation of the source. The generalization of both the Reissner-Nordstr\"om and Kerr solutions, known as Kerr-Newman (KN) metric \cite{newman1965note,newman1965metric} describes the spacetime of a rotating charged source. Unlike the RN case, the electromagnetic source is not pointlike but a distribution of mass, charge, and current on a disk \cite{israel1970source, pekeris1987electromagnetic}. The Kerr-Newman  metric in Boyer-Lindquist coordinates \cite{boyer1967maximal} is given by \cite{newman1965metric}
\begin{equation}
    \begin{aligned}
g=&-\frac{\Delta}{\rho^{2}}\left[a \sin ^{2}(\theta) d \varphi-d u\right]^{2}\\
&+ \frac{\rho^{2}}{\Delta} d r^{2}+\rho^{2} d \theta^{2}+\frac{\sin ^{2} \theta}{\rho^{2}}\left[\left(r^{2}+a^{2}\right) d \varphi-a d u\right]^{2} 
\end{aligned} \label{KNmetric}
\end{equation}
where
\begin{equation}
    \Delta(r)=r^{2}-2 M r+a^{2}+Q^{2}
\end{equation}
and
\begin{equation}
    \rho^{2}(r, \theta)=r^{2}+a^{2} \cos ^{2} \theta ,
\end{equation}
for a source of  angular momentum  per unit mass $a$, mass $M$ and charge $Q$.  The coordinates  $(u,r,\theta,\varphi)$ are Schwarzschild-like coordinates.
 {The quantities $r$, $\rho$, $\sqrt\Delta$, $u$ and $a$ all have the dimensions of lengths.}
The metric \eqref{KNmetric} reduces to the Kerr metric for $Q=0$ and to the Reissner-Nordstr\"om metric  
\eqref{metric} for $a=0$. If $a=Q=0$ one gets the Schwarzschild metric.

In the Minkowski coframe, given by
\begin{eqnarray}
&&\df e^0=-\frac{\sqrt\Delta}{\rho}(du-a\sin^2\theta d\varphi)\\
&&\df e^1=\frac{\rho}{\sqrt\Delta}dr\\ &&\df e^2=\rho d\theta\\
&&\df e^3=\frac{\sin\theta}{\rho}(a du-(r^2+a^2)d\varphi)
\end{eqnarray} 
the 1-form electromagnetic potential  in the Kerr-Newman spacetime is given by
$\df A=A_0\df e^0$ which, in local coordinates reads as \cite{israel1970source}
\begin{equation}
   \df  A=  {\frac{Qr}{\rho\sqrt\Delta}\df e^0=}-
    \frac{Q r}{\rho^{2}}\left(d u-a \sin ^{2} \theta d \varphi\right).
\end{equation}
From the Faraday 2-form $\df F=d\df A= \df E\wedge du +\df B$, it follows that the electric field $1$-form is given by 
\begin{equation}
   \df E= \frac{Q}{\rho^4}\left[r^2 -a^2 \cos^2 \theta   \right]dr -\frac{Q}{\rho^4}a^2  \sin 2\theta \, (r d\theta ) \label{E_KN}
\end{equation}
and the magnetic flux density $2$-form by 
\begin{eqnarray}
 \df   B= &  & \frac{Q}{r\rho^4}\left[r^2 -a^2 \cos^2 \theta   \right]a \sin\theta \,  (r\sin\theta d \varphi) \wedge dr \nonumber \\ 
   &  + & \frac{2Q}{r\rho^4}a  (r^2 + a^2)\cos\theta \, (r d\theta)\wedge (r\sin\theta d\varphi). \label{B_KN}
\end{eqnarray}
 {In the Minkowski coframe, the expression of $\df F$ takes a simpler form
\begin{equation}
    \df F=-\frac{Q(\rho^2-2r^2)}{\rho^4}\df e^0\wedge\df e^1-\frac{2Qar\cos\theta}{\rho^4}\df e^2\wedge e^3
\end{equation}
and the evaluation of the Hodge star is made easier because of fortunate simplifications:
\begin{eqnarray}
    \star_4\df F &=& \sqrt{-\eta}F^{ab}\epsilon_{abcd}\df e^c\wedge \df e^d\nonumber\\
    &=& \frac{Q(\rho^2-2r^2)}{\rho^4}\df e^2\wedge \df e^3-\frac{2Qar\cos\theta}{\rho^4}\df e^0\wedge \df e^1.
\end{eqnarray}
This result is finally reverted to the local coframe:
\begin{eqnarray}
    \star_4\df F &=&{\frac{Qa}{\rho^4}\Bigl[\Bigl(
    -(\rho^2-2r^2)\sin\theta  d\theta+r\cos\theta dr\Bigr)}\wedge du\nonumber\\
    &&+(\rho^2-2r^2))a^{-1}d\theta\wedge d\varphi
    +2ar\sin\theta\cos\theta d\varphi\wedge dr\Bigl]
\end{eqnarray}
This leads to}
 $\df D$ and $\df H$ fields from the Maxwell 2-form $\df G=\star_4 \df F=\df D - \df H\wedge du$. We get for the electric flux density
\begin{eqnarray}
 \df   D  = & & \frac{Q}{r^2\rho^4} \left[  r^2 -a^2 \cos^2 \theta  \right](r^2 + a^2)\, (r d\theta ) \wedge (r\sin\theta  d\varphi) \nonumber
   \\
    & - & \frac{Q}{\rho^4}a^2\sin 2\theta \, (r\sin\theta \, d\varphi) \wedge dr . \label{D_KN}
\end{eqnarray}
Comparing \eqref{D_KN} to \eqref{E_KN} we see that $E_{\theta} = D_{r \varphi }$ but $E_r \neq D_{\theta\varphi}$. 

The magnetic field is given by 
\begin{eqnarray}
  \df  H = & &\frac{Q}{r\rho^4} {\left[\rho^2-2r^2  \right]}a \sin\theta \, (r d\theta) 
  + \frac{2Q}{\rho^4}{a r} \cos\theta dr \label{H_KN}
\end{eqnarray}
and we have that
$H_{\theta}=B_{r\varphi}$ but $H_r \neq B_{\theta\varphi}$, from Eqs. \eqref{B_KN} and \eqref{H_KN}. To summarize, the vacuum response in the KN spacetime is anisotropic: polarization and magnetization occur only for the radial components of the respective fields.

In order to visualize the classical vacuum polarization effects of the KN spacetime  we map the electromagnetic fields into Minkowski spacetime by assuming that Eqs. \eqref{E_KN}-\eqref{H_KN} are in a flat background in spherical coordinates. Using this approach, we show in Figs. \ref{electric_lines} and \ref{magnetic_lines} the electromagnetic field lines in the $z-x$ plane. They may be compared to the  plots of $\textbf{E}$ and $\textbf{H}$ presented in \cite{pekeris1987electromagnetic}, keeping in mind that they are plotted in a different background. While we forced the field lines to be in Minkowski spacetime, Ref. \cite{pekeris1987electromagnetic} plots the fields in the KN curved background described in a Cartesian system (that is asymptotically flat) introduced originally by Kerr \cite{kerr1963gravitational}. It is clear from the plots that the electromagnetic sources are located on a disk on the equatorial plane. Both charge and current densities are singular  on the disk rim, in agreement with Refs. \cite{israel1970source,pekeris1987electromagnetic}.  Furthermore, the polarization is relevant only in the region near the sources. Away from the sources, $\textbf{E} \sim \textbf{D}$ and $\textbf{B} \sim \textbf{H}$.

\begin{figure}[htp!]
    \centering 
      \includegraphics[width=0.49\columnwidth]{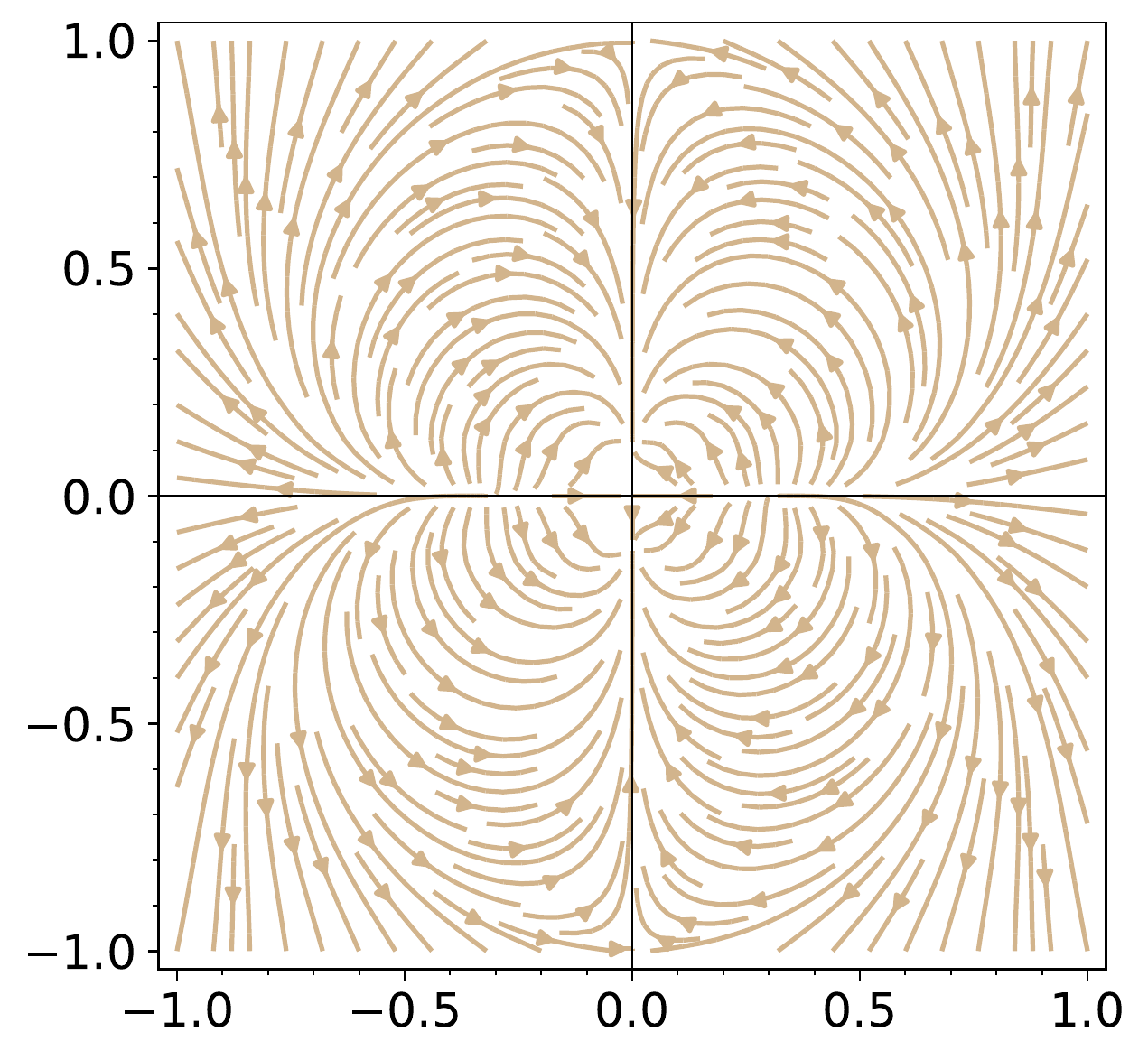} 
      \includegraphics[width=0.49\columnwidth]{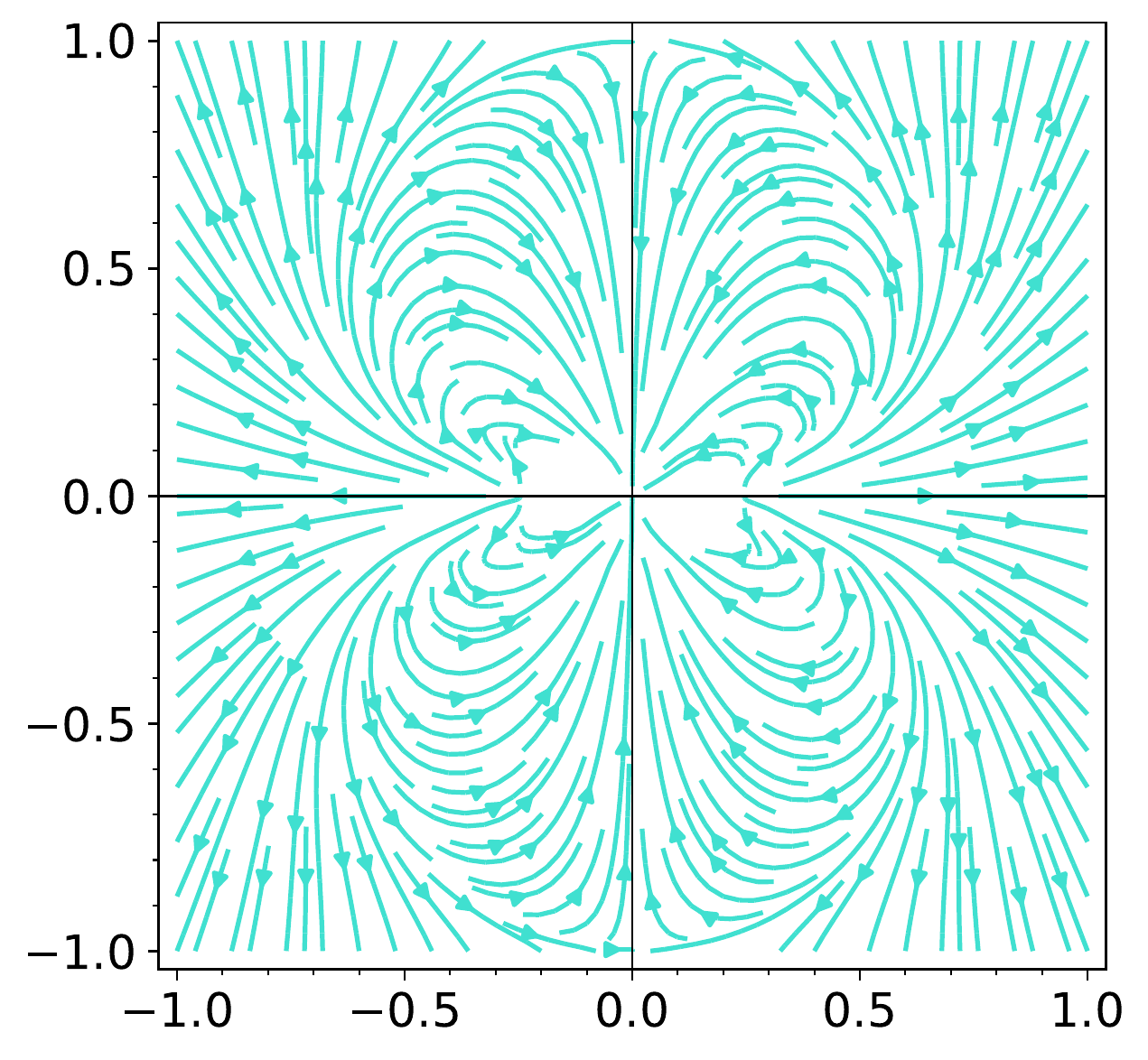} 
    \caption{Kerr-Newman electric field $\mathbf{E}$ (left) and electric flux density $\mathbf{D}$ (right) lines. The scale was set by choosing $a=Q=1$.}
   \label{electric_lines}
\end{figure}
\begin{figure}[htp!]
    \centering 
      \includegraphics[width=0.49\columnwidth]{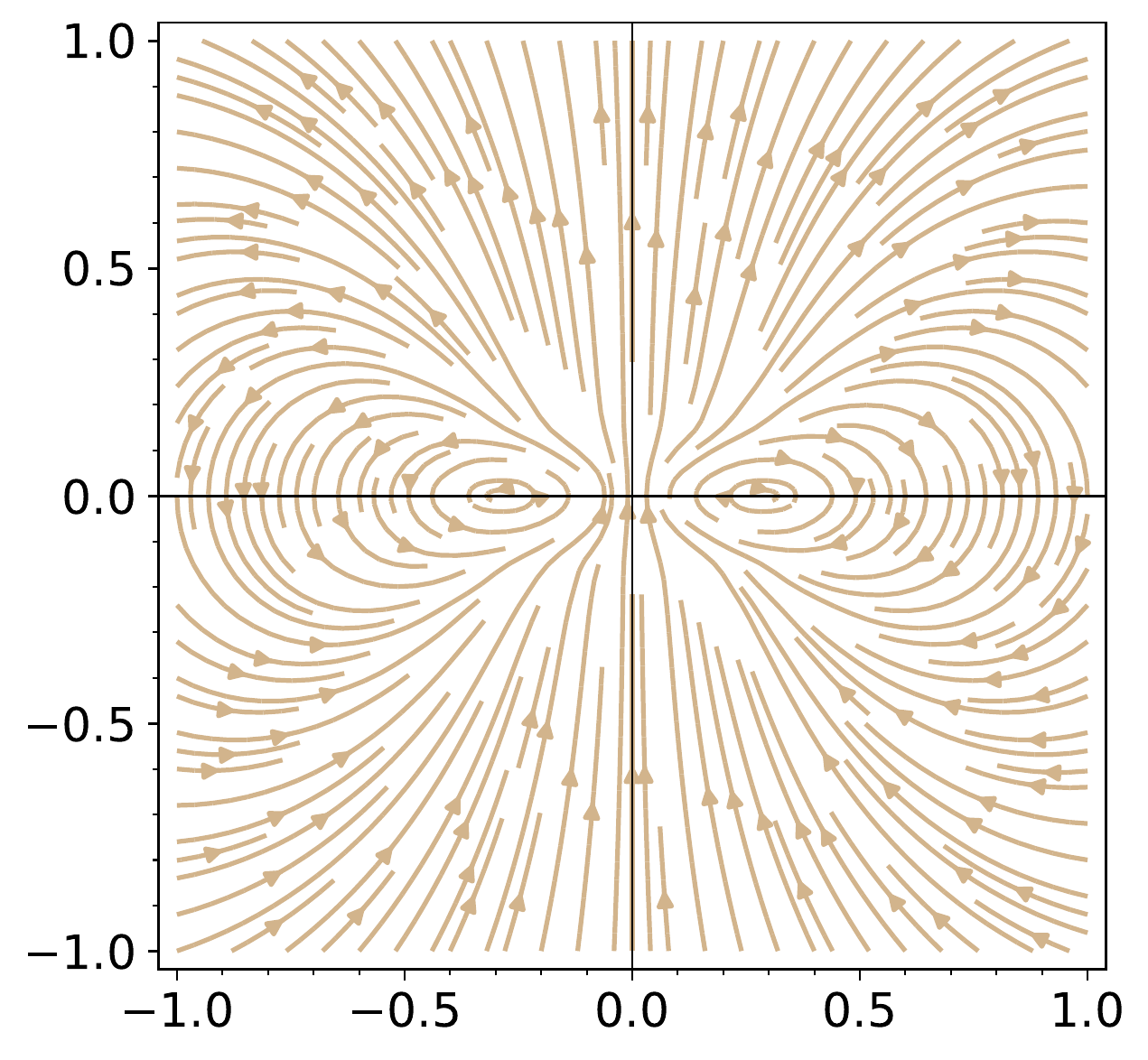} 
      \includegraphics[width=0.49\columnwidth]{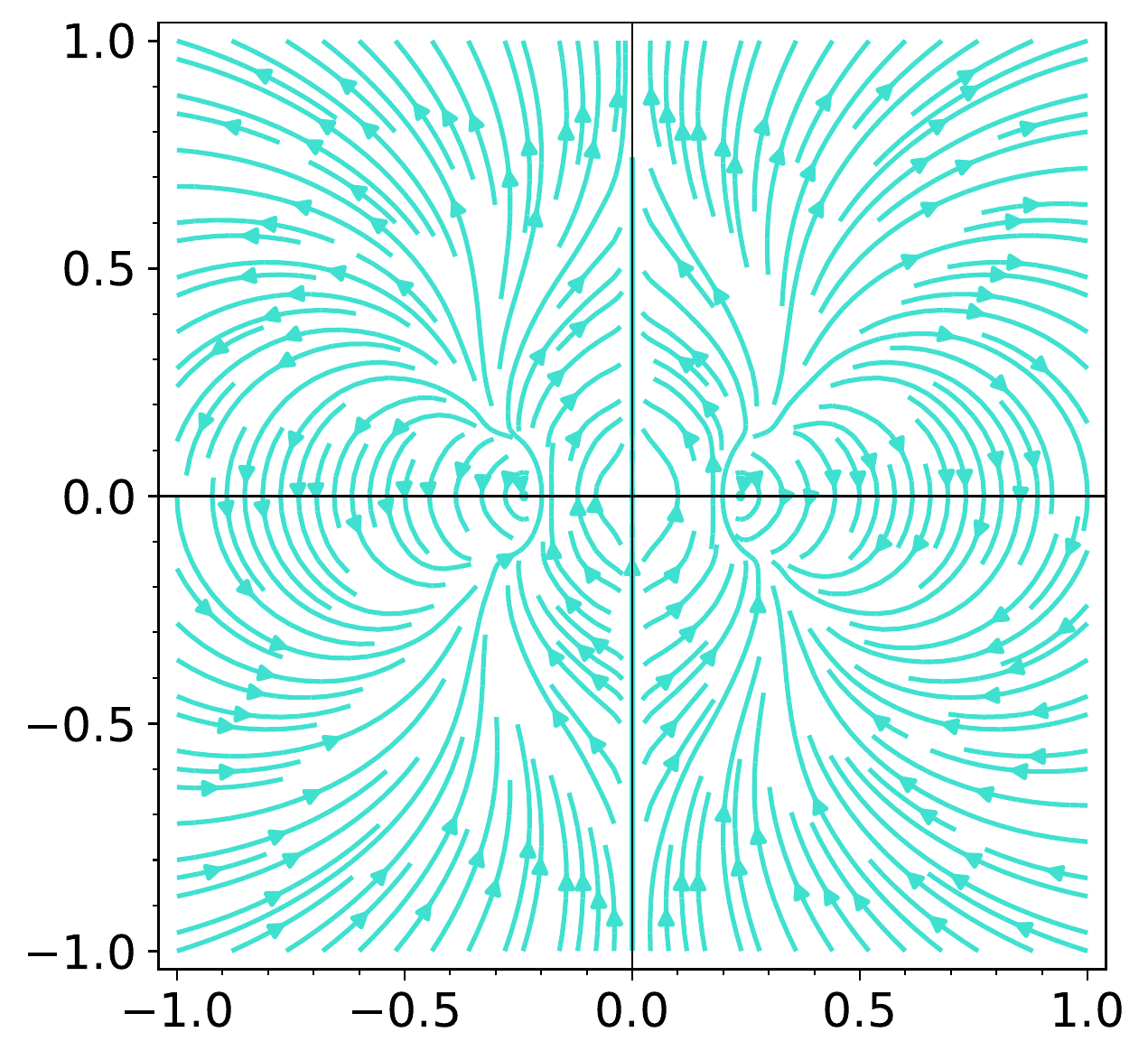} 
    \caption{Kerr-Newman magnetic flux density $\mathbf{B}$
 (left) and magnetic field $\mathbf{H}$ (right) lines. The scale was set by choosing $a=Q=1$.}
   \label{magnetic_lines}
\end{figure}

\section{Conclusions \label{conclsection}}

The examples studied in this article show that both classical vacuum polarization and magnetization do occur in some curved spacetimes with electromagnetic sources. \revision{These properties may be useful for a possible detection of non trivial spacetime geometries from distant observations.}

\revision{Mathematically,  the non vanishing polarisation/magnetization} is a consequence of the Hodge star operation that incorporates the spacetime geometry  while providing the transformation between the ``fundamental'' ($E$, $B$) and ``excitation''  ($D$, $H$) fields.  
\revision{To understand this property, we can compare the standard approach of electrodynamics in terms of tensors with the one privileged in this paper in terms of differential forms. In terms of tensors, one can sum up the basic postulates of electrodynamics to Bianchi equation for the Faraday tensor  $\partial_\lambda F_{\mu\nu}+\partial_\mu F_{\nu\lambda}+\partial_\nu F_{\lambda\mu}=0$,
and to the equation of motion for the Maxwell tensor $\partial_\mu G^{\mu\nu}=-\sqrt{-g}J^\nu$. At this point, there is still missing an action to provide a relation between the two second-rank tensors $F_{\mu\nu}$ and $G^{\mu\nu}$. In Maxwell
electrodynamics, this action is  
 $S[A_\sigma]=
\int \sqrt {-g}\ \!d^4x\ \! \bigl({\textstyle\frac 14} F_{\mu\nu}F^{\mu\nu}-A_\mu J^\mu\bigr)$, the minimisation of which leads to the equations of motion $\frac 1{\sqrt {-g}}\partial_\mu\bigl(\sqrt {-g}F^{\mu\nu}\bigr)=-J^\nu$.
In terms of differential forms, Bianchi identity reads as $d\df F=0$ and the equation of motion reduces to $d\df G=-\star_4\df J$ and, again, the system of equations is not complete in absence of a relation between $\df G$ and $\df F$. This relation is either provided again by an action as in equation (\ref{Eq-action}), of directly by the duality relation $\df G=\star_4\df F$ called constitutive relation. If $\df G$ was obtained independently of the coordinate system via generalized Gauss or Amp\`ere theorems, the metric dependence of the Hodge star operation makes $\df F$ depend on the local coordinates and possibly lead to a non trivial polarizability/permeability in these coordinates.
}

We remark that our study encompasses static fields and therefore is not equivalent to the well-known interpretation of curved spacetime as an effective anisotropic medium for light propagation \cite{Plebanski60}. Moreover, in all cases studied here, both gravitational and electromagnetic fields are solutions of the coupled Einstein-Maxwell equations, \revision{since the electromagnetic field's energy-momentum tensor generates a very weak gravitational field, thus its contribution to gravity in the presence of matter sources is usually neglected: this is in particular the case for a distant observer who might detect the imprint of the polarization/magnetization in the geometry. A possible extension of the results presented here may consist in describing the full backreaction of the electromagnetic field via the energy-momentum tensor \cite{tsagas2004electromagnetic}.}

Analogies between gravity and elasticity of continuum media have been explored by many authors (see for instance \cite{tartaglia2010tensor,millette2013elastodynamics,bohmer2014continuum}). 
Let us quote for example Landau and Lifshitz\cite{Landau:1975pou}:
\begin{quote}
    ``We may say that with respect to its effect on the electromagnetic field a static gravitational field plays the role of a medium with electric and magnetic permeabilites.''
\end{quote} 
On the other hand, the coupling of electromagnetic fields to  elastic deformations gives rise to well-known phenomena like piezoelectricity  (piezomagnetism) and the not-so-well-known flexoelectricity (flexomagnetism) \cite{bardzokas2007mathematical}. Further, in higher dimensional gravity, electroelastic effects have been found in strained charged branes  \cite{armas2012black,armas2013electroelasticity}. This leads us to conclude that the results  obtained here for the vacuum dielectric and magnetic response functions, with the associated gravitational fields, may perhaps be realized in electro-magneto-elastic media as analog models for Einstein-Maxwell solutions.  Conversely, one may propose an Einstein-Maxwell approach to electro-magneto-elastic materials. \revision{Indeed, elasticity in continuum mechanics has been long related to gravity \cite{AndreiDSakharov_1991}, their similarity is made explicit when the equation obeyed by the deformation field of an elastic medium is written as an "Einstein equation", as shown in \cite{padmanabhan2004gravity}. This naturally allows for a generalization of the elastic Einstein equation to the elastic Einstein-Maxwell equations by considering electromagnetic fields in material media and their coupling to elasticity. In other words: the rewriting of the electro-magneto-elastic equations as effective Einstein-Maxwell equations.
}

The classical response of vacuum may provide new tools for the astronomical search of cosmic strings and wormholes. The former are expected to produce observable signatures such as gravitational lensing \cite{morganson2010direct}, anisotropic patterns in Cosmic Microwave Background, Kaiser-Stebbins effect \cite{Planck2013,jeong2005search} or powerful bursts of gravitational waves due to strings cusps \cite{aasi2014constraints}. Wormholes could be observed from lensing effects \cite{cramer1995natural}-\cite{abe2010gravitational} or from the iron line spectrum of their accretions disk \cite{zhou2016search}. In this work, we showed that the vacuum constitutive relations depend on the position and on the string/wormhole parameters. This leaves a usable imprint on the propagating waves: as is known from Rytov law, the polarization plane of a wave is indeed likely to rotate when propagating inside inhomogeneous media \cite{bliokh2004modified}. 

\section*{Acknowledgements}
FM thanks Conselho Nacional de Desenvolvimento Científico e Tecnológico (CNPq),  for partially supporting this work.

\revision{
\section*{Appendix}
The Hodge dual operator $\star_n$ is an invertible linear map between any $p$-form $\df v\in\Lambda^p\left(\mathcal{M}\right)$ and its dual ($n-p$)-form $\star_n\df v \in \Lambda^{n-p}\left(\mathcal{M}\right)$ such that \cite{Baez}
\begin{equation}
\df u\wedge\left(\star\df v\right)=\left\langle \df u,\df v \right\rangle \sqrt{\left|\text{det}g_{ab}\right|}\: dx^1\wedge..\wedge dx^n \label{Hodge-def-gen}
\end{equation}
(here $\df u$ is of the same degree as $\df v$) with $n$ the dimension of the manifold $\mathcal M$, here $n=4$. {In the case considered here, 
\begin{equation}
\df u\wedge\left(\star_4\df v\right)=\left\langle \df u,\df v \right\rangle  r^2\sin\theta \:dt\wedge dr\wedge d\theta \wedge d\varphi .\label{Hodge-def}
\end{equation}}
Again, $\df u$ is of the same degree as $\df v$ and the inner product $\left\langle\  ,\ \right\rangle$ between two \textit{p}-forms obeys:
\begin{eqnarray}
&&p=1:\;\;\left\langle dx^{\mu}, dx^{\nu} \right\rangle=g^{\mu\nu} \\
&&p>1:\;\;\left\langle dx^{\mu_1}\wedge..\wedge dx^{\mu_p}, dx^{\nu_1}\wedge..\wedge dx^{\nu_p} \right\rangle \nonumber \\
&&\;\;\;\;\;\;\;\;\;\;\;\;\;\;\;=\left|\begin{pmatrix}
g^{\mu_1 \nu_1} & \cdots & g^{\mu_1 \nu_p}\\
\vdots&\ddots&\vdots\\
g^{\mu_p \nu_1}&\cdots&g^{\mu_p \nu_p}
\end{pmatrix}\right|.
\end{eqnarray}
A useful property is that in a Minkowski frame, the Hodge dual of the Faraday form is simply
\be  \star_4\df F = \sqrt{-\eta}F^{ab}\epsilon_{abcd}\df e^c\wedge \df e^d\ee
with $-\eta=1$ and $\df e^a$ the Minkowski cotetrads.
}

\section*{References}
\bibliography{references}

\providecommand{\newblock}{}
\begin{thebibliography}{10}
\expandafter\ifx\csname url\endcsname\relax
  \def\url#1{{\tt #1}}\fi
\expandafter\ifx\csname urlprefix\endcsname\relax\def\urlprefix{URL }\fi
\providecommand{\eprint}[2][]{\url{#2}}

\bibitem{dirac1934discussion}
Dirac P~A 1934 Discussion of the infinite distribution of electrons in the
  theory of the positron {\em Mathematical Proceedings of the Cambridge
  Philosophical Society\/} vol~30 (Cambridge University Press) pp 150--163

\bibitem{Furry1934theory}
Furry W~H and Oppenheimer J~R 1934 {\em Physical Review\/} {\bf 45} 245

\bibitem{heisenberg1989bemerkungen1}
Heisenberg W 1934 {\em Zeitschrift f{\"u}r Physik\/} {\bf 90} 209--231
  \urlprefix\url{https://doi.org/10.1007/BF01333516}

\bibitem{uehling1935polarization}
Uehling E~A 1935 {\em Physical Review\/} {\bf 48} 55

\bibitem{weisskopf1936electrodynamics}
Weisskopf V 1936 {\em Kgl. Dan. Vid. Selsk. Mazt-fys. Medd\/} {\bf 24} 3--39

\bibitem{leuchs2020qed}
Leuchs G, Hawton M and S{\'a}nchez-Soto L~L 2020 {\em Physics\/} {\bf 2} 14--21

\bibitem{cabral2017electrodynamics}
Cabral F and Lobo F~S 2017 {\em Foundations of Physics\/} {\bf 47} 208--228

\bibitem{cabral2017electrodynamicsII}
Cabral F and Lobo F~S 2017 {\em The European Physical Journal Plus\/} {\bf 132}
  1--16

\bibitem{PhysRevD.105.105026}
Berche B, Fumeron S and Moraes F 2022 {\em Phys. Rev. D\/} {\bf 105}(10) 105026
  \urlprefix\url{https://link.aps.org/doi/10.1103/PhysRevD.105.105026}

\bibitem{info}
 {\em In this paper we use the metric signature $-,+,+,+$ and the convention
  $\epsilon_{0123}=1$ for the Levi-Civita symbol.\/}

\bibitem{fumeron2020improving}
Fumeron S, Berche B and Moraes F 2020 {\em American Journal of Physics\/} {\bf
  88} 1083--1093

\bibitem{Schw1916}
Schwarzschild K 1916 {\em (translation and foreword by S.Antoci and A.Loinger),
  arXiv:physics/9905030\/}

\bibitem{Bekenstein}
Bekenstein J~D 1971 {\em Physical Review D\/} {\bf 4} 2185

\bibitem{Vilenkin:2000jqa}
Vilenkin A and Shellard E~P~S 2000 {\em {Cosmic Strings and Other Topological
  Defects}\/} (Cambridge University Press) ISBN 978-0-521-65476-0

\bibitem{FB2023}
Fumeron S and Berche B 2023 {\em Eur. Phys. J. Spec. Top., arXiv:2209.07743\/}
  \urlprefix\url{http://dx.doi.org/10.1140/epjs/s11734-023-00803-x}

\bibitem{Planck2013}
Ade P~A~R, Aghanim N, Armitage-Caplan C, Arnaud M, Ashdown M, Atrio-Barandela
  F, Aumont J, Baccigalupi C, Banday A~J and et~al 2014 {\em Astronomy \&
  Astrophysics\/} {\bf 571} A25 ISSN 1432-0746
  \urlprefix\url{http://dx.doi.org/10.1051/0004-6361/201321621}

\bibitem{auclair2023window}
Auclair P, Leyde K and Steer D~A 2023 {\em Journal of Cosmology and
  Astroparticle Physics\/} {\bf 2023} 005

\bibitem{Linet1999}
Linet B 1999 {\em Classical and Quantum Gravity\/} {\bf 16} 2947--2953
  \urlprefix\url{https://doi.org/10.1088%2F0264-9381%2F16%2F9%2F312}

\bibitem{kim2001exact}
Kim S~W and Lee H 2001 {\em Physical Review D\/} {\bf 63} 064014

\bibitem{morris1988wormholes}
Morris M~S and Thorne K~S 1988 {\em American Journal of Physics\/} {\bf 56}
  395--412

\bibitem{garcia2002zero}
Garcia N, Ponizovskaya E and Xiao J~Q 2002 {\em Applied physics letters\/} {\bf
  80} 1120--1122

\bibitem{schwartz2003total}
Schwartz B~T and Piestun R 2003 {\em JOSA B\/} {\bf 20} 2448--2453

\bibitem{liu2017manipulating}
Liu S, Zhou J, Han Y, Yu X, Chen H and Lin Z 2017 Manipulating electromagnetic
  waves with zero index materials {\em Wave Propagation Concepts for
  Near-Future Telecommunication Systems\/} (IntechOpen)

\bibitem{melvin1964pure}
Melvin M~A 1964 {\em Physics Letters\/} {\bf 8} 65--68

\bibitem{melvin1965dynamics}
Melvin M 1965 {\em Physical Review\/} {\bf 139} B225

\bibitem{lim2018electric}
Lim Y~K 2018 {\em Physical Review D\/} {\bf 98} 084022

\bibitem{ernst1976black}
Ernst F~J 1976 {\em Journal of Mathematical Physics\/} {\bf 17} 54--56

\bibitem{Bytsenko2003}
Bytsenko A and Goncharov Y 2003 {\em International Journal of Modern Physics
  A\/} {\bf 18} 2153--2157

\bibitem{footnoteBytsenko}
 {\em {\rm We disagree here with Ref. \cite{Bytsenko2003} where we believe that
  the derivation erroneously assumes that $\Lambda$ is a constant.}\/}

\bibitem{kerr1963gravitational}
Kerr R~P 1963 {\em Physical Review Letters\/} {\bf 11} 237

\bibitem{newman1965note}
Newman E~T and Janis A 1965 {\em Journal of Mathematical Physics\/} {\bf 6}
  915--917

\bibitem{newman1965metric}
Newman E~T, Couch E, Chinnapared K, Exton A, Prakash A and Torrence R 1965 {\em
  Journal of mathematical physics\/} {\bf 6} 918--919

\bibitem{israel1970source}
Israel W 1970 {\em Physical Review D\/} {\bf 2} 641

\bibitem{pekeris1987electromagnetic}
Pekeris C and Frankowski K 1987 {\em Physical Review A\/} {\bf 36} 5118

\bibitem{boyer1967maximal}
Boyer R~H and Lindquist R~W 1967 {\em Journal of mathematical physics\/} {\bf
  8} 265--281

\bibitem{Plebanski60}
Plebanski J 1960 {\em Physical Review\/} {\bf 118} 1396

\bibitem{tsagas2004electromagnetic}
Tsagas C~G 2004 {\em Classical and Quantum Gravity\/} {\bf 22} 393

\bibitem{tartaglia2010tensor}
Tartaglia A and Radicella N 2010 {\em Classical and Quantum Gravity\/} {\bf 27}
  035001

\bibitem{millette2013elastodynamics}
Millette P~A 2013 {\em Progress in Physics\/} {\bf 1} 55--59

\bibitem{bohmer2014continuum}
B{\"o}hmer C~G and Downes R~J 2014 {\em International Journal of Modern Physics
  D\/} {\bf 23} 1442015

\bibitem{Landau:1975pou}
Landau L~D and Lifschits E~M 1975 {\em {The Classical Theory of Fields}\/}
  ({\em Course of Theoretical Physics\/} vol Volume 2) (Oxford: Pergamon Press)
  ISBN 978-0-08-018176-9

\bibitem{bardzokas2007mathematical}
Bardzokas D~I, Filshtinsky M~L and Filshtinsky L~A 2007 {\em Mathematical
  methods in electro-magneto-elasticity\/} vol~32 (Springer Science \& Business
  Media)

\bibitem{armas2012black}
Armas J, Gath J and Obers N~A 2012 {\em Physical Review Letters\/} {\bf 109}
  241101

\bibitem{armas2013electroelasticity}
Armas J, Gath J and Obers N~A 2013 {\em Journal of High Energy Physics\/} {\bf
  2013} 35

\bibitem{AndreiDSakharov_1991}
Sakharov A~D 1991 {\em Soviet Physics Uspekhi\/} {\bf 34} 394
  \urlprefix\url{https://dx.doi.org/10.1070/PU1991v034n05ABEH002498}

\bibitem{padmanabhan2004gravity}
Padmanabhan T 2004 {\em International Journal of Modern Physics D\/} {\bf 13}
  2293--2298

\bibitem{morganson2010direct}
Morganson E, Marshall P, Treu T, Schrabback T and Blandford R~D 2010 {\em
  Monthly Notices of the Royal Astronomical Society\/} {\bf 406} 2452--2472

\bibitem{jeong2005search}
Jeong E and Smoot G~F 2005 {\em The Astrophysical Journal\/} {\bf 624} 21

\bibitem{aasi2014constraints}
Aasi J, Abadie J, Abbott B, Abbott R, Abbott T, Abernathy M, Accadia T,
  Acernese F, Adams C, Adams T {\em et~al.\/} 2014 {\em Physical Review
  Letters\/} {\bf 112} 131101

\bibitem{cramer1995natural}
Cramer J~G, Forward R~L, Morris M~S, Visser M, Benford G and Landis G~A 1995
  {\em Physical Review D\/} {\bf 51} 3117

\bibitem{abe2010gravitational}
Abe F 2010 {\em The Astrophysical Journal\/} {\bf 725} 787

\bibitem{zhou2016search}
Zhou M, Cardenas-Avendano A, Bambi C, Kleihaus B and Kunz J 2016 {\em Physical
  Review D\/} {\bf 94} 024036

\bibitem{bliokh2004modified}
Bliokh K~Y and Bliokh Y~P 2004 {\em Physical Review E\/} {\bf 70} 026605

\bibitem{Baez}
Baez J and Muniain J~P 1994 {\em Gauge fields, knots and gravity\/} vol~4
  (World Scientific Publishing Company, Singapore)

\end{thebibliography}

\end{document}